\documentclass{afig} 

\usepackage[utf8]{inputenc} 
\usepackage[english]{babel} 
\usepackage[T1]{fontenc}
\usepackage{subcaption}
\usepackage{float}

\ifpdf \usepackage[pdftex]{graphicx} \pdfcompresslevel=9
\else \usepackage[dvips]{graphicx} \fi

\title[DrEAM: a New metaphor for UAV Control]{DrEAM: a New Exocentric Metaphor for Complex Path Following to Control a UAV Using Mixed Reality}

\author[B. Wojtkowski, P. Castillo et I. Thouvenin]%
       {B. Wojtkowski$^1$
        et P. Castillo$^1$ 
				et I. Thouvenin$^1$
        \\
				 $^1$Laboratoire Heudiasyc, CNRS UMR 7253, Sorbonne Universités, \\ Université de Technologie de Compiègne,60200 Compiègne, France
       }

\begin{document}

\maketitle

\begin{figure*}[h!]
	\centering
	\includegraphics{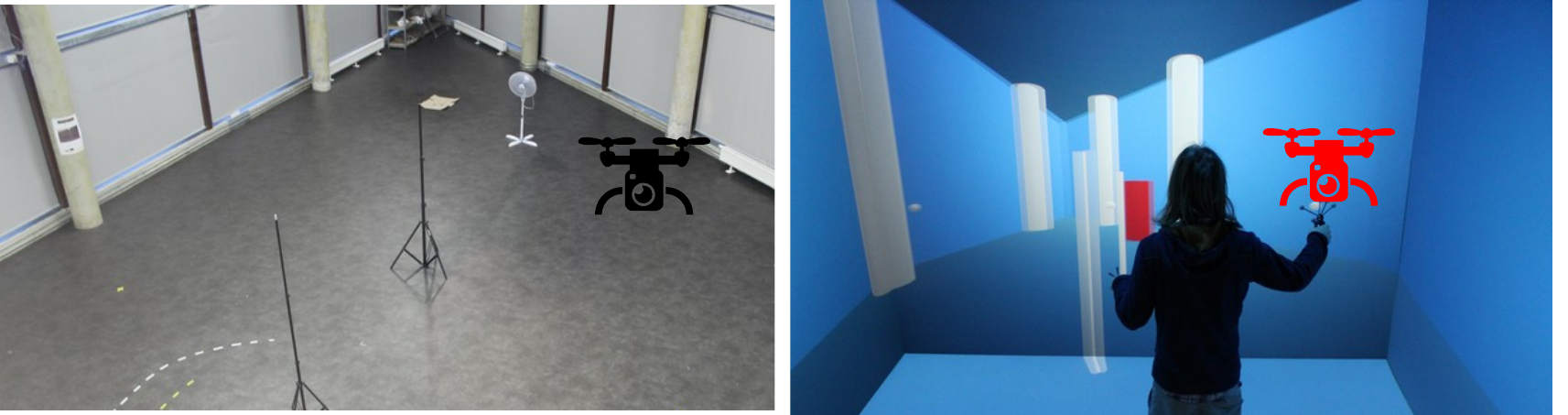}
	\caption{DrEAM is a new Metaphor for UAV control : user has a virtual UAV in his hand (left side) and controls a real UAV (right side)}
\end{figure*}

\begin{abstract}
Teleoperation of Unmanned Aerial Vehicles (UAVs) has recently become an noteworthly research topic in the field of human robot interaction.
Each year, a variety of devices is being studied to design adapted interface for diverse purpose such as view taking, search and rescue operation or suveillance. New interfaces have to be precise, simple and intuitive even for complex path planning. Moreover, when teleoperation involves long distance control, user needs to get proper feedbacks and avoid motion sickness.
In order to overcome all these challenges, a new interaction metaphor named DrEAM (Drone Exocentric Advanced Metaphor) was designed. User can see the UAV he is controlling in a virtual environment mapped to the real world. He can interact with it as a simple object in a classical virtual world.
An experiment was lead in order to evaluate the perfomances of this metaphor, comparing performance of novice user using either a direct-view joystick control or using DrEAM.

\end{abstract}

\keywords{UAV control, Mixed Reality for Robotic, Exocentric Metaphor}
\section{Introduction} 
Teleoperation of UAVs has become a widespread topic covering various sectors such as view taking, merchandise delivery, or surveillance and rescue missions. Most of UAV flight can be fully automated such as in transports field, but a lot of missions still need to be performed by humans .
Therefore it has become necessary to design the most efficient interface possible in order to control the UAV easily and precisely in miscellaneous situations including those requiring complex path following and high understanding of UAV's environment.
There are usualy two main approaches \cite{fernandez_natural_2016}: Natural User Interfaces (NUI) and Graphical User Interfaces (GUI), coupled with joystick controls. First tend to be easier to master while second tend to be more precise. A third class called Brain Computer Interface (BCI) is being developped but with very few success concerning UAV control at the moment.
Over last years, a lot of interfaces were developped in particular for direct view control and the purpose of this paper is to draw inspiration from these metaphors and propose one for UAV control over long distance.
In this paper, we investigate a new interaction metaphor supposed to increase the ease of control over a complex path.
We propose a teleoperation based on an exocentric metaphor (DrEAM) to overcome motion sickness and allow user to have precise movement in an inertial frame. DrEAM uses a World In Miniature (WIM) modeling of the world in a CAVE-like environment.
Since direct view control with joystick is one of the most widespread and efficient control at the moment, the main idea is to create a metaphor that has at least same performances as this control.
After a short look at the existing metaphors for UAV control, DrEAM will be introduced in details. Then experiment will be detailed in particular method and results optained.

\section{Related Works}
\subsection{NUI for UAV control}
Research in UAV teleoperation has resulted in a diversity of control devices for varied purposes. For some applications such as live-action technique or domestic use, user usually see directly the controlled UAV.
For more technical application such as safety, security or rescue purpose, it becomes necessary to have non direct view mataphor and to model the environment.

\subsection{NUI for direct view control}
Natural Interaction have already been investigated and a lot of modalities have been explored such as hand tracking using Leap Motion Controller \cite{daniel_liebeskind_leap_2013}\cite{gubcsi_ergonomic_2018}, head tracking \cite{pfeil_exploring_2013} or basically gesture tracking \cite{sanna_kinect-based_2013} 
A few Natural User Interfaces (NUI) for UAV control have already been compared and classified  \cite{peshkova_natural_2017},\cite{pfeil_exploring_2013}. Interaction metaphors can be grouped in class which rely on different mental models. In the imitative class the teleoperated object imitates user's behaviour (for example if the user move his hand towerd left, UAV flies toward left). In instrumented class, the teleoperated object is controlled through an third party item (which usually doesn't realy exist), for example as if one were flying a kite or as if one had a true joystick in the hand.
Interaction vocabulary of an interaction must be as simple as possible, close to natural behavior, coherent and culture dependant \cite{peshkova_natural_2017}.
A trend is to mix interaction modalities \cite{fernandez_natural_2016}\cite{ng_collocated_2011} such as voice and hand control. 

\subsection{Joystick-Based Control and GUI}
Even with all these devices, direct control of UAV is still mainly accomplished by using joystick-based interfaces. They use to be related with a Graphical User Interface (GUI) providing informations such as the speeds, thrusts or battery level. They allow users to perform precise movements and to have robust control of the teleoperated UAV. Problem of such interface concerns mainly the expertise needed to perform precisely complex path.

\subsection{Non-direct view UAV Control}
In Virtual Reality, interaction with immersive environment use to be classified in two categories\cite{jung_review_2014} : egocentric metaphores and exocentric metaphores.
On one hand, egocentric metaphors allow user to be embedded in the world and perform actions inside it. On the other hand, exocentric metaphors allow user to stay outside the world and to have multiple points of view. 
In the field of teleoperation, most of the metaphors in immersive environments are egocentric metaphors and user controls the UAV from its point of view with a camera feedback \cite{prexl_user_2017}\cite{miehlbradt_data-driven_2018}.
The main issue of such interaction metaphors is that they suppose the user to watch continuously a camera stream, which usually implies a high motion sickness due to movements, in particular using head mounted device (HMD) \cite{chen_human_2007}. \\

\section{DrEAM}

\subsection{Studied metaphor}

\begin{figure}
	\centering
	\begin{subfigure}[t]{\linewidth}
		\centering
		\includegraphics[width=\textwidth]{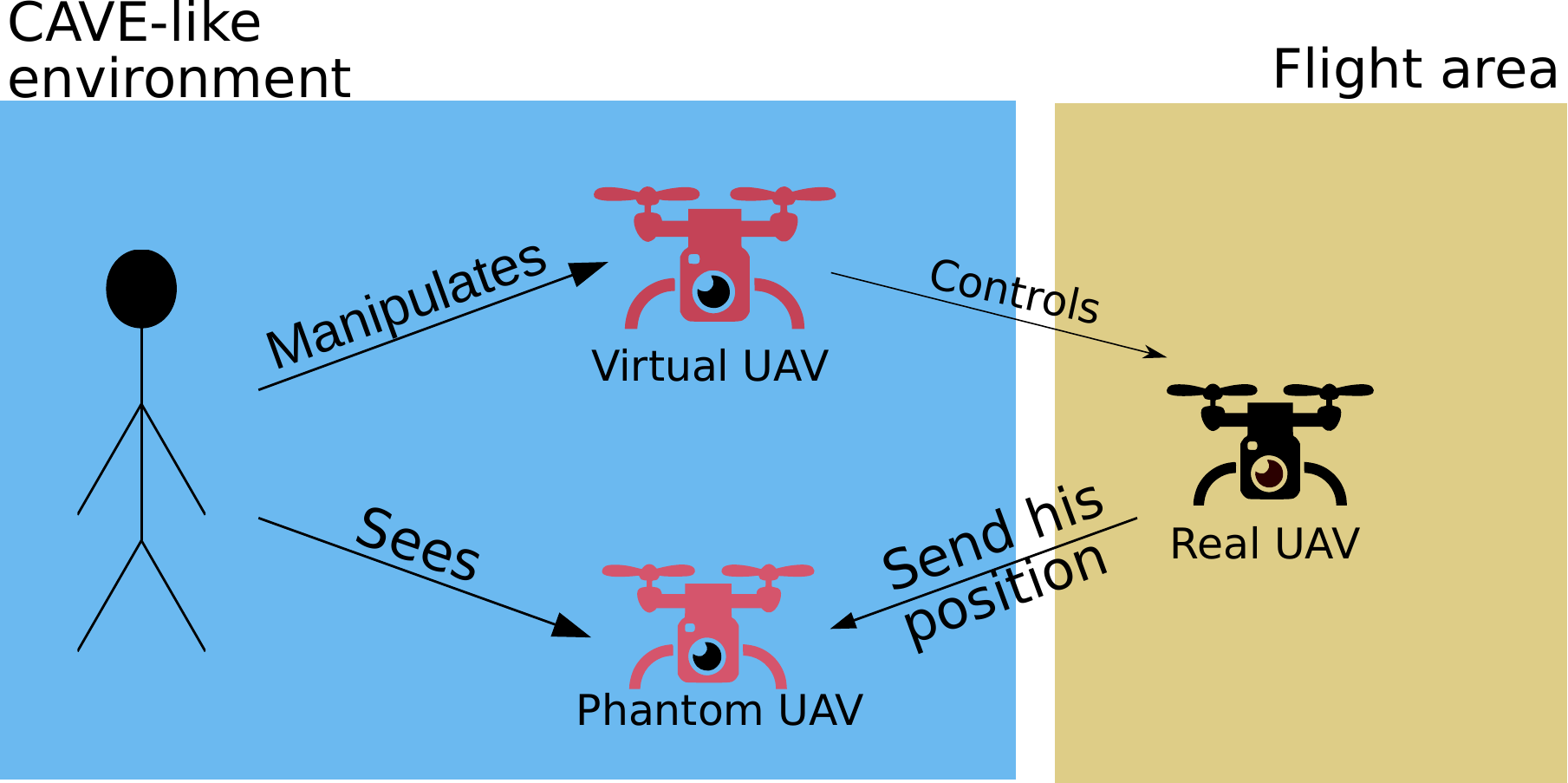}
		\caption{Presentation of a classical joystick}
	\end{subfigure}

	\begin{subfigure}[t]{\linewidth}
		\centering
		\includegraphics[width=\textwidth]{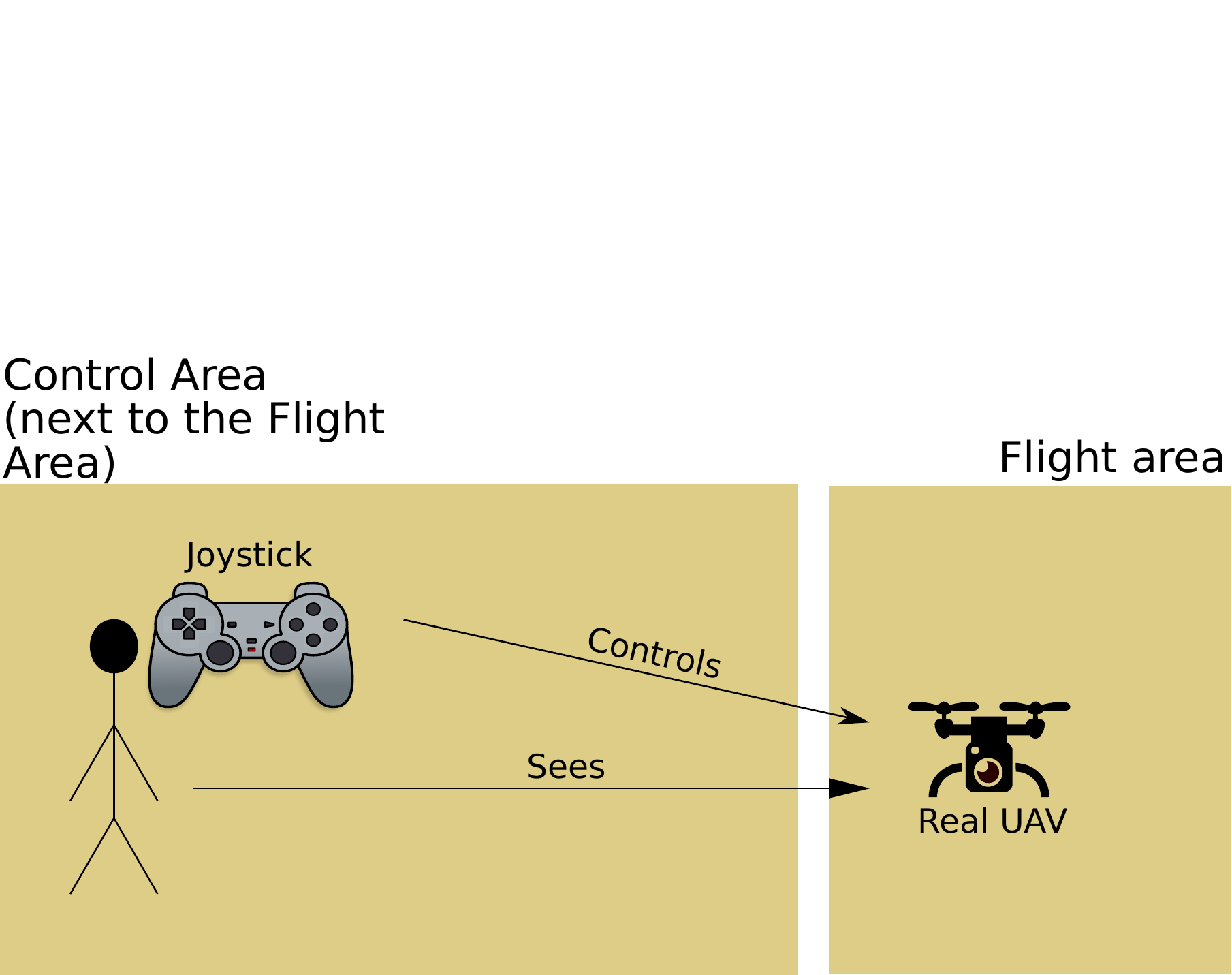}
		\caption{Presentation of DrEAM}
	\end{subfigure}
	\caption{Comparizon of both metaphors}
\end{figure}

With DrEAM, user is spectator of the virtual environment. He can see a 3D reproduction of the world  move inside and rotate the environment using basic commands.
The Virtual UAV (VUAV) is an object of the virtual world that user can take in his hand by pressing a specific button and release by releasing it. While the user takes the VUAV, he can rotate and translate it simply by moving his hand, as if the UAVs where really in his hand. The real UAV (RUAV), in the flight area, has then to follow the path given by the user through VUAV. User can see a second UAV (the Phantom UAV, PUAV) in the virtual world, that represents the feedback of the real UAV. This UAV is placed at the last known position and orientation of the real UAV. See figure \ref{metaphore} for more informations and a comparison with joystick control.
\subsection{Metaphor details}
\subsubsection{Manipulating the leader UAV}


At the beginning of the flight, the VUAV is placed at the position corresponding to the RUAV.
The VUAV can be manipulated by the user by pressing the "Take" button on the index of the wand. The UAV can be taken only if user's hand is in his hit box (which has the same size as the object) in order to have a precise control of the UAV. User can see visual feedback informing him that any object of the world can be taken or is taken.
The VUAV can be oriented and translated as the user pleases and the RUAV will be placed at the corresponding place and orientation in the real world. 
.
Color of the VUAV changes depending on its state (can be taken, cannot be taken, is taken).


\subsubsection{Sensitive feedback}

The RUAV is represented to the user through two different feedback. User can see the position and orientation of the UAV looking at the PUAV position. He also has an indication of the speed of the UAV with a sound feedback.

\subsection{Technical details}
\subsubsection{Hardware Details}
Embedded part of the system consists in a Parrot AR-UAV 2 (capable of doing stationnary flight) flying in an inner flight area and tracked by an Opti-Track system. A CAVE-Like environment was used, also using Opti-Track to track stereo  vision glasses and a PS Wand.
Both communicate together and with their respective VRPN server in a common wireless network. Both platforms have their respective switch and both switch are bound toward a bigger network.

\subsubsection{Software Details}
On the UAV, an embedded Linux runs a fl-AIR (framework libre AIR) application. Fl-AIR is a C++ based framework for UAV control. On the CAVE-like platform, a Unity plugin named TransOne encapsulates data from VRPN into Unity Objects for a better use in Unity.

\subsubsection{Representation of the key points and modelling} 
The whole environment was modeled in DrEAM, where start and arrival points were represented by little spheres, and the target by a big red cylinder. In this case, the subject could only see UAV through the feedback of our application.




\section{Method}
\subsection{Evaluation of metaphors for UAV control}
A few tests were already lead by researchers in order to test the advantages of such UAVs teleoperation devices. These tests generally consist in a comparison between a given metaphore and a few other interface performing a simple control task.
Researchers usually base their analysis on quantitative data such as completion time, number of errors, precision \cite{pfeil_exploring_2013}\cite{yu_human-robot_2014}, in some studies, they also check the speed variations or the quantity of movements performed by users or even by the UAV \cite{prexl_user_2017}, and also on qualitative subjective informations using TLX forms \cite{yu_human-robot_2014} or other specific forms \cite{gubcsi_ergonomic_2018}.
TLX forms evaluates ergonomics of the control metapher. User evaluates his performance among 6 criterias: Mental demand, Physical Demand, Temporal Demand (this depicts the stress involved by the control task), performance (this depicts the sensation of success), Effort, Frustration (this depicts the sensation of UAV's obedience).

\subsection{Hypothesis}
In order to test the advantages of DrEAM over direct view control, an experimental study was lead on eight volunteers testing following hypothesis: \textit{DrEAM increases the control ergonomy for unexperimented users, without precision loss, in comparison with a joystick control, in particular concerning movements composing degrees of freedom.}
We made following null-hypothesis\\
\begin{itemize}
\item \textbf{H0} DrEAM has no impact on Mental demand for an unexperimented user performing a complex task 
\item \textbf{H1} DrEAM has no impact on Physical demand for an unexperimented user performing a complex task 
\item \textbf{H2} DrEAM has no impact on Temporal demand for an unexperimented user performing a complex task  
\item \textbf{H3} DrEAM has no impact on subjective Performance for an unexperimented user performing a complex task  
\item \textbf{H4} DrEAM has no impact on Effort for an unexperimented user performing a complex task  
\item \textbf{H5} DrEAM has no impact on Frustration for an unexperimented user performing a complex task  
\item \textbf{H6} User of dream have better performances while accomplishing a complex control task
\end{itemize}
\subsection{Task}
A protocol was designed to test these hypothesis. Tested device was DrEAM in comparison with joystick control. For each of these devices, participants had to train four minutes on a specific task, then had three minutes to complete it as much as possible and as precisely as possible.

\begin{figure}[htb!]
	\center
	\includegraphics[width=\linewidth]{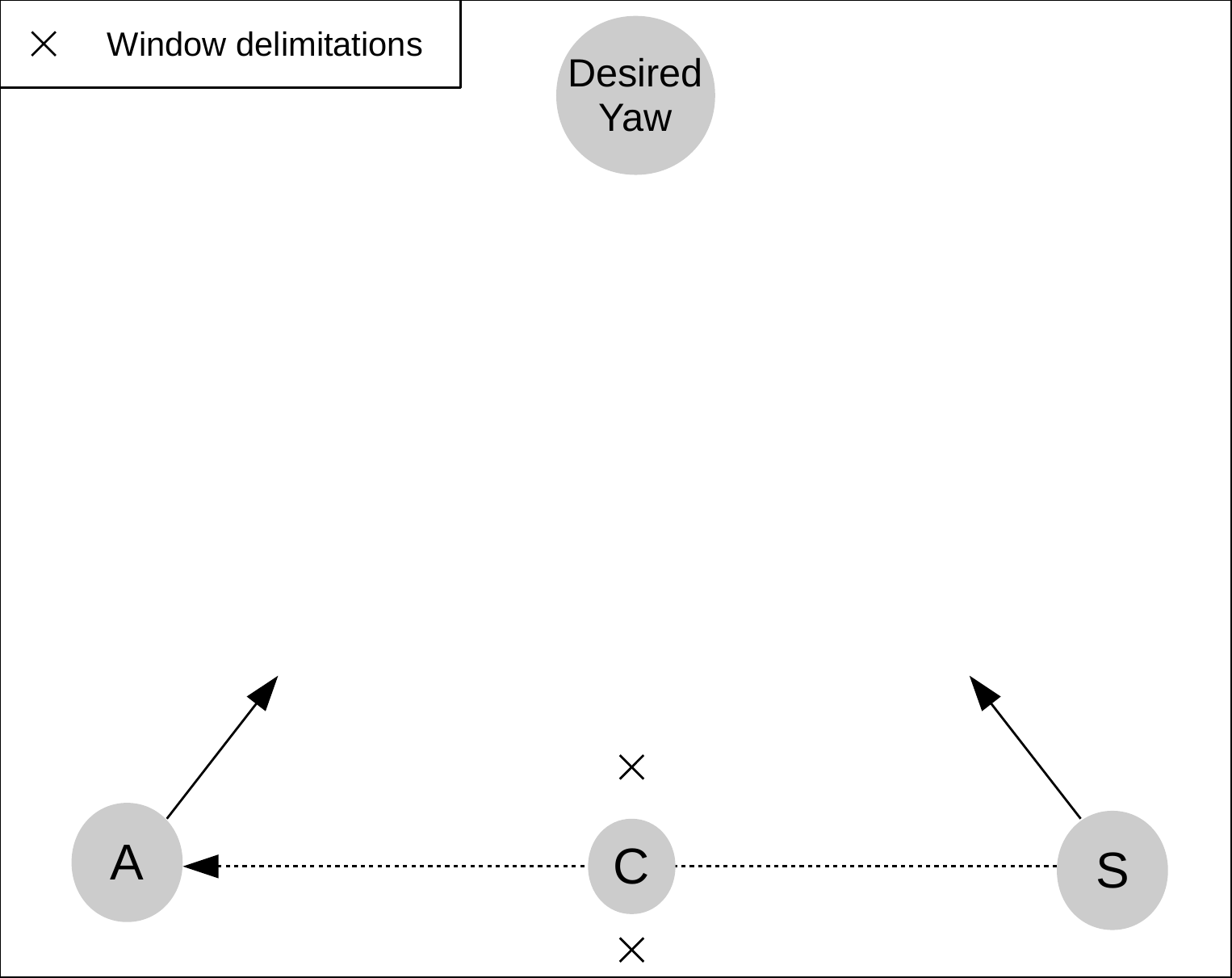}
	\caption{Navigation Task \\ (UAV is supposed to go from start (S) to arrival (A) through checkpoint (C), with head pointing toward desired yaw)}
	\label{path}
\end{figure}

Each participant had to follow the navigation task described in figure \ref{path}. The UAV was brought at the start point at the beginning of the experiment by an experimenter. Then, the user had to move laterally and regulate the yaw in order to have the UAV head always pointing the target. Moreover, user had to lead the UAV at a checkpoint located at the center of a window (depicted with two tripods). Each participant had three instructions:
\begin{enumerate}
\item UAV must not be put in danger 
\item Task must be accomplished as precisely as possible 
\item Task must be accomplished as quick as possible
\end{enumerate}

Every flight information were logged in order to analyze them after flight and determine interesting information such as errors to a reference path, or completion time. 
After each part of the experiment, a NASA-TLX form was given to the participant.

\begin{figure}[htb!]
	\includegraphics[height=3in, width=\linewidth]{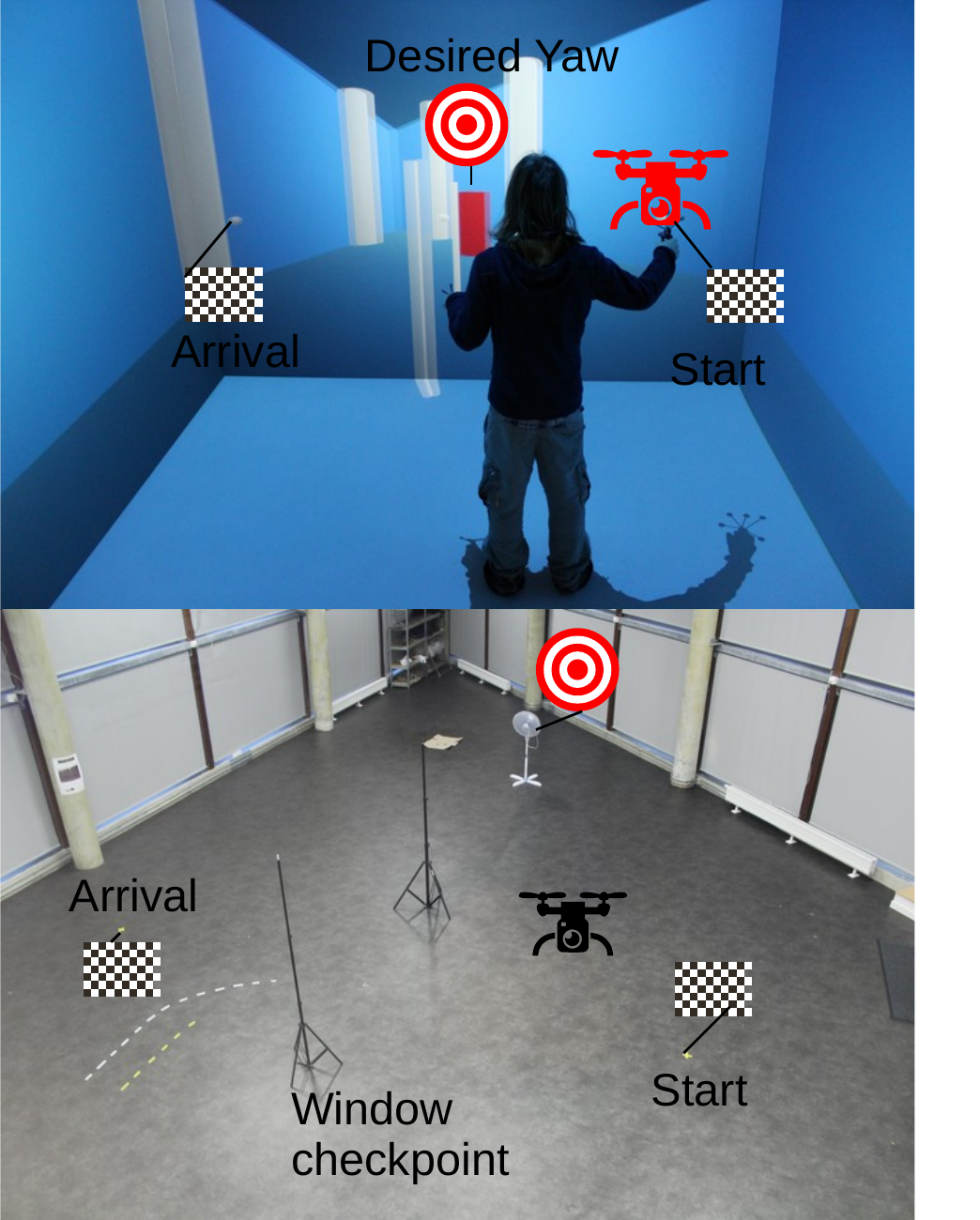}
	\caption{Experimental setup}
	\label{schema}
\end{figure}

\subsubsection{Controls}
For direct view piloting, the UAV is directly piloted using fl-AIR with a PID using position.
Controls with the joystick were standards: user could control the horizontal translations in the body frame with the left joystick and the yaw with right joystick. In this case, user could see the UAV directly.
The task is supposed to be complicated for the pilot because of frames used for joystick control (see figure \ref{frames})

\subsubsection{Representation of the keypoints in the flight zone and perspective effects}
To neutralize most of the perspective effects, UAV was set to fly close to the ground (about 1 meter), where points reported on figure \ref{path} were represented by cross. Moreover, two tripods were placed at the center of the path to depict a fight window. This was supposed to give the user physical marker for the depth.

\subsection{Experiment groups}
Each participant was given a preliminary form in order to assign him an experiment group. Groups were balanced according to six factors: sex, gender, age, experience in virtual worlds, experience in immersive virtual world and experience in control of UAV. \\
Each participant performed a specific task using either a joystick or DrEAM and then performed it a second time using the other device, each experiment group had a specific running order.

\subsection{Protocol details}
For each passage, the participant was briefed on controls, teaching him how to translate and rotate the UAV. 
Then they were brought in the control zone (virtual or real) and the experimenter showed them the task, moving in the flight area as the UAV should, so that the user could not misinterpret the task.
After this short introduction, the experimenter had the UAV take off and go at the start point, then the participant had four minutes to train. After this time, the experimenter had the UAV land, he changed the battery and had the UAV take off again and started logging flight information. Then the user had three minutes to perform the task following the instructions.
At the end of each task, participant was given a NASA-TLX form.

\subsection{Data collected}
\begin{figure}[t]
	\center
	\includegraphics[width=\linewidth]{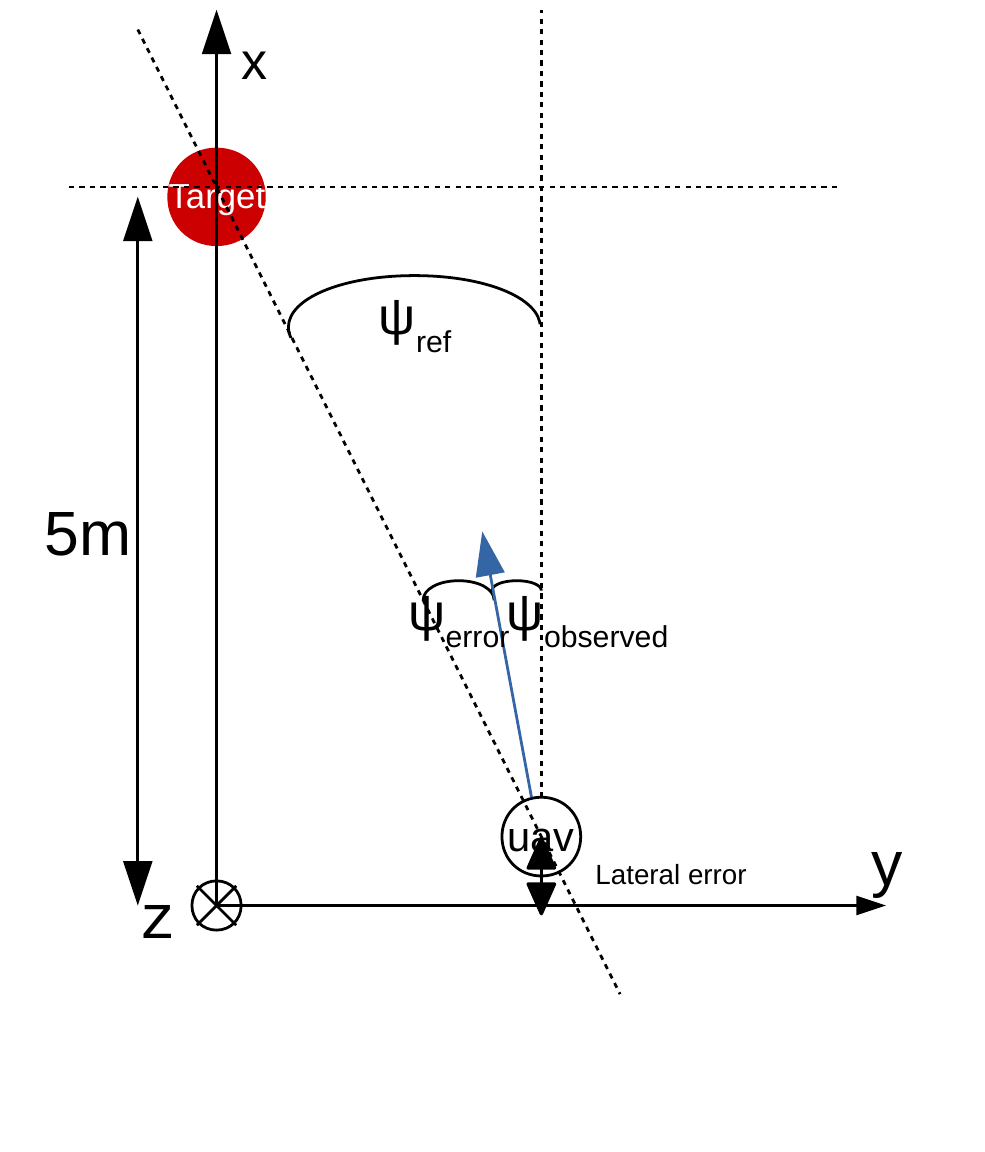}
	\caption{Calculated errors for results analysis}
	\label{measuredValues}
\end{figure}

In addition to NASA-TLX forms, we logged every useful flight information. Flight logs consisted in a set positions with corresponding timestamp, orientation, speeds and thrusts. We calculated some perfomance indicators from these data.\\
Indicators were mean lateral error (MLE) to the theoretical path, the mean completion time (MCT) of a journey and the mean yaw error (MYE) to the theoretical path.
For each completed journey, the lateral error is the average orthogonal distance to the path of all point logged by the UAV, in fact value of the x-axis (see figure \ref{measuredValues}), the MLE is the mean of all these data for all journeys.\\

For the MYE, the process is the same.
For a given point with an y-axis value equal to y and x-axis equal to x, theoretical yaw is given by $\psi_{ref}=\arctan(y/(5-x))$, yaw error is $\psi_{error} = |\psi_{observed}-\psi_{ref}|$, 
For a given journey, the completion time is the time between the end of the stop at the start point and the full stop at the arrival point, MCT is the mean completion time for all journeys.

\section{Results}

\begin{figure}[htb!]
	\centering
	\begin{subfigure}[t]{\linewidth}
		\centering
		\includegraphics[width=\textwidth, height=2.5in]{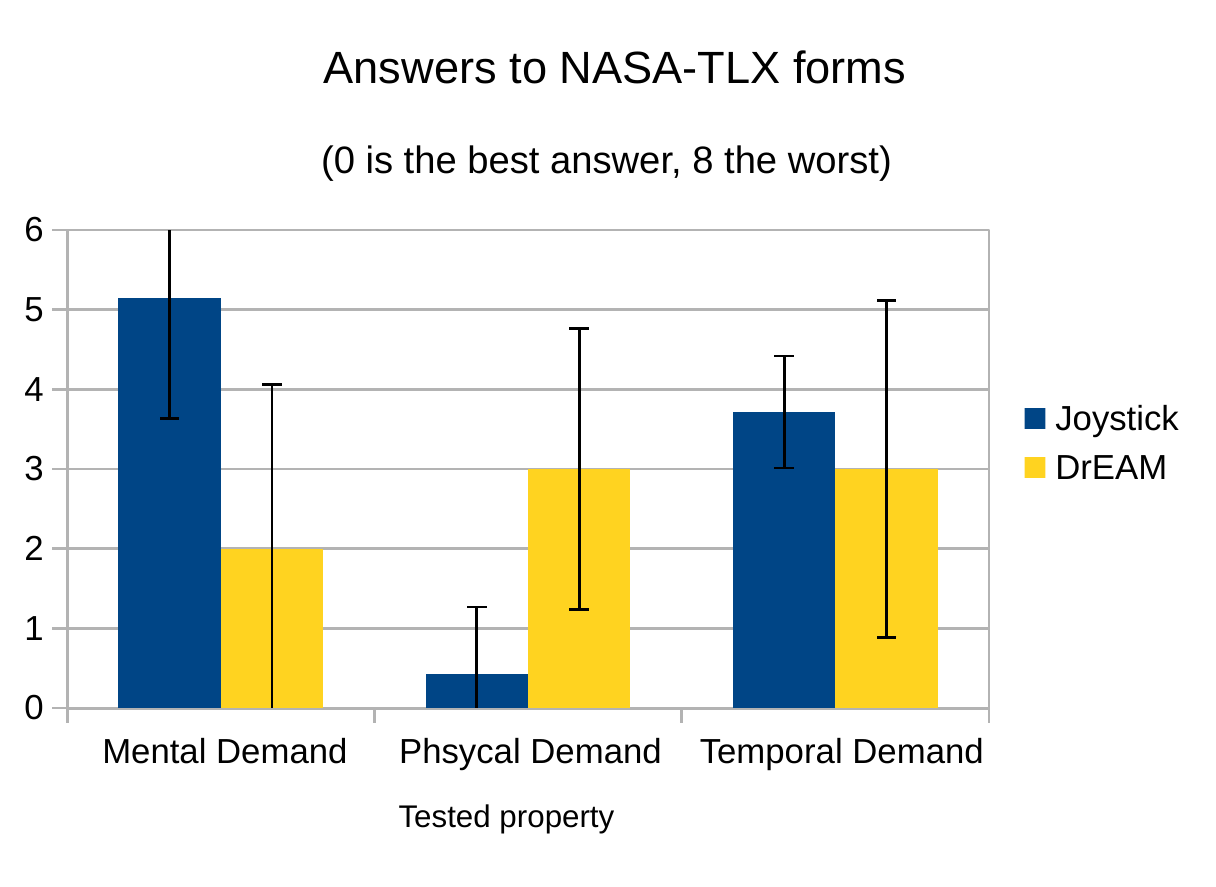}
	\end{subfigure}%

	\begin{subfigure}[t]{\linewidth}
		\centering
		\includegraphics[width=\textwidth, height=2.5in]{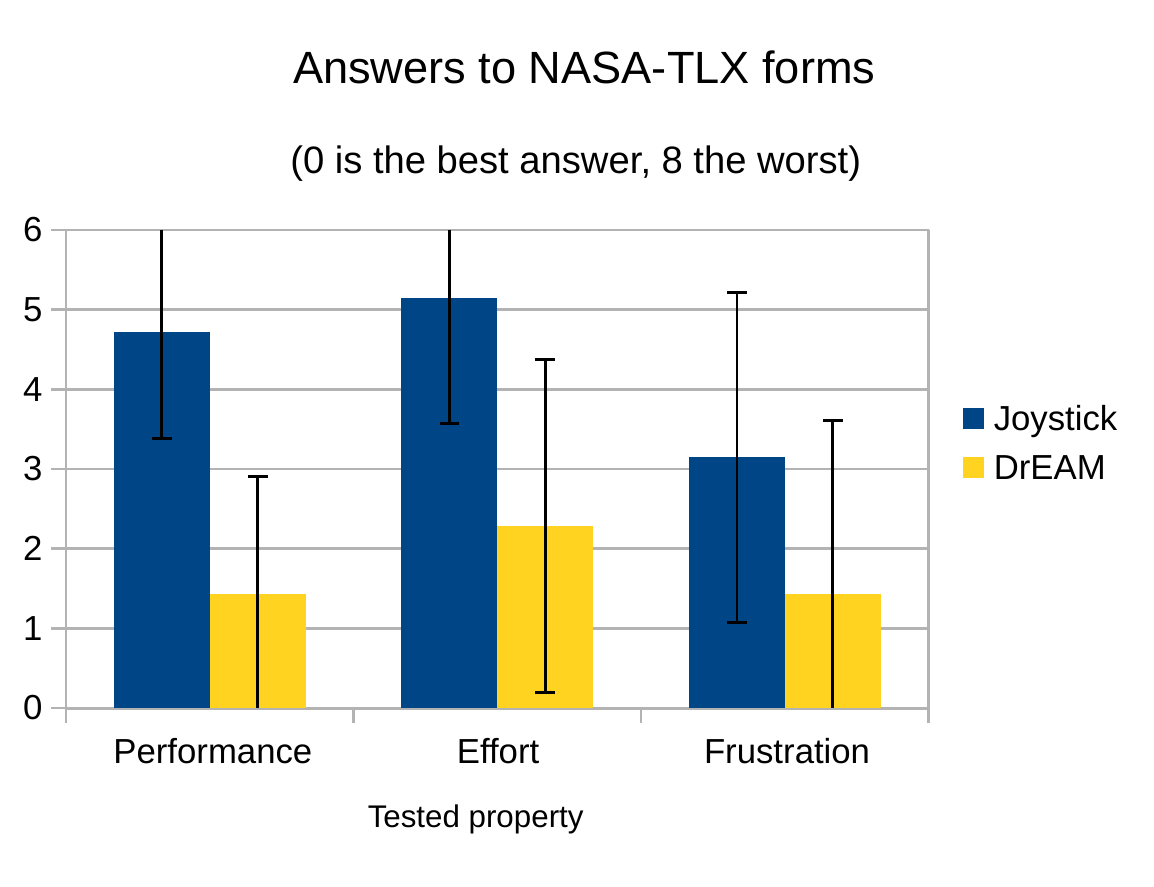}
	\end{subfigure}
	\caption{Results of the TLX-Questionnaire}
	\label{resTLX}
\end{figure}

A total of eight participants volunteered for our study, most of subjects were men and had never piloted a UAV.
NASA-TLX results are summed up in figure \ref{resTLX}. 

Student tests were applied for each hypothesis from the NASA-TLX, since we only tested two modalities for the control interface. Results of the tests are consigned in table \ref{res-Student}
\begin{table}[htb]
	\centering
  \begin{tabular}{|l|c|r|}
		\hline
		Hypothesis & t-Value & $H_0$ Result  \\
		\hline \hline
		Mental Demand & 0.013 & Rejected \\ \hline
		Physical Demand & 0.007 & Rejected \\ \hline
		Temporal Demand & 0.269 & Not Rejected \\ \hline
		Performance & 0.003 & Rejected \\ \hline
		Effort & 0.022 & Rejected \\ \hline
		Frustration & 0.14 & Not Rejected \\ \hline
	\end{tabular}
	\caption{t-Value for each hypothesis (for each line, $H_0$ is: "DrEAM has no impact on X")}
	\label{res-Student}
\end{table}

\begin{figure}[htb!]
	\centering
	\begin{subfigure}[t]{\linewidth}
		\centering
		\includegraphics[width=\textwidth, height=2in]{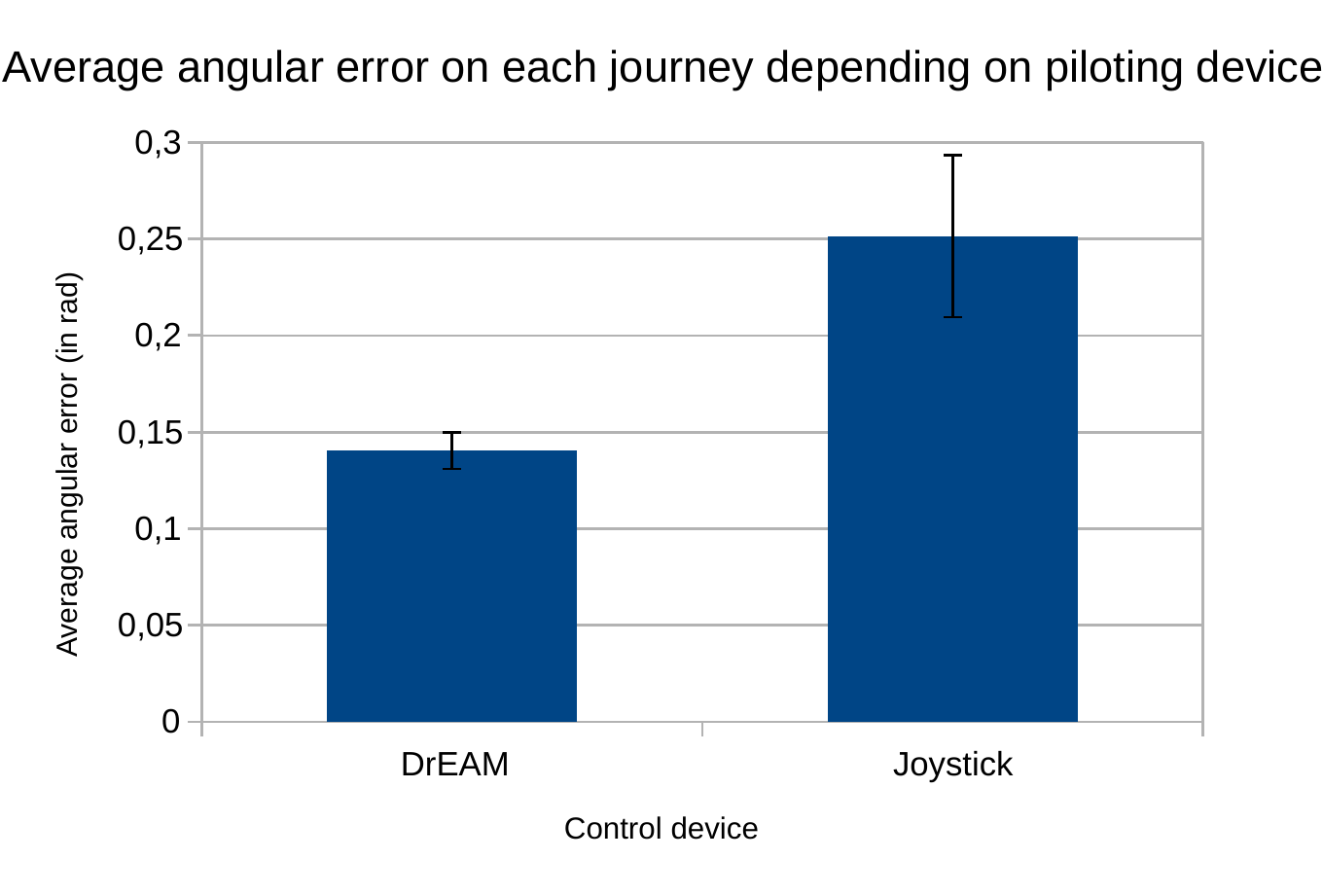}
		\caption{MYE over all journeys}
	\end{subfigure}%

	\begin{subfigure}[t]{\linewidth}
		\centering
		\includegraphics[width=\textwidth, height=2in]{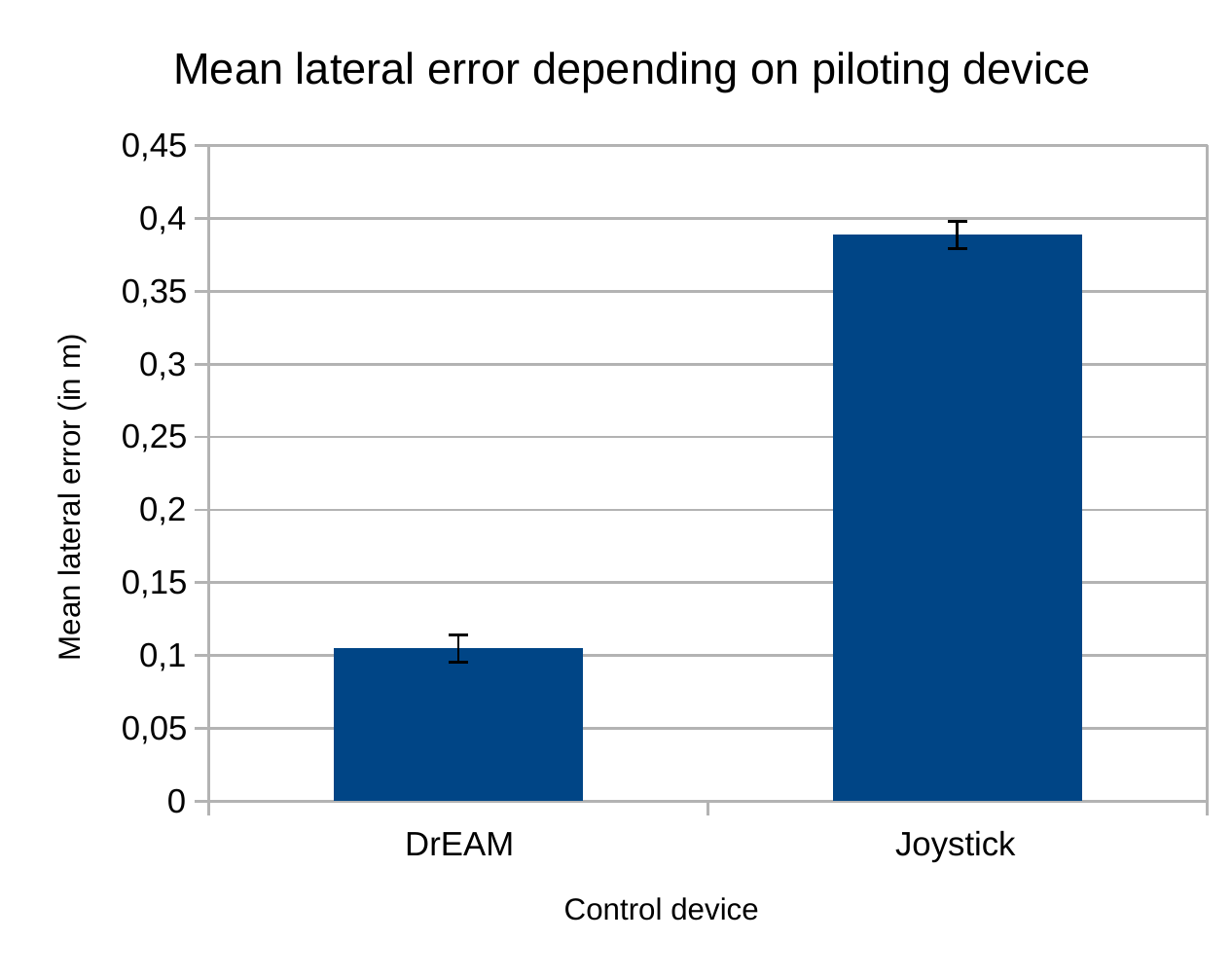}
		\caption{MLE over all journeys}
	\end{subfigure}

	\begin{subfigure}[t]{\linewidth}
		\centering
		\includegraphics[width=\textwidth, height=2in]{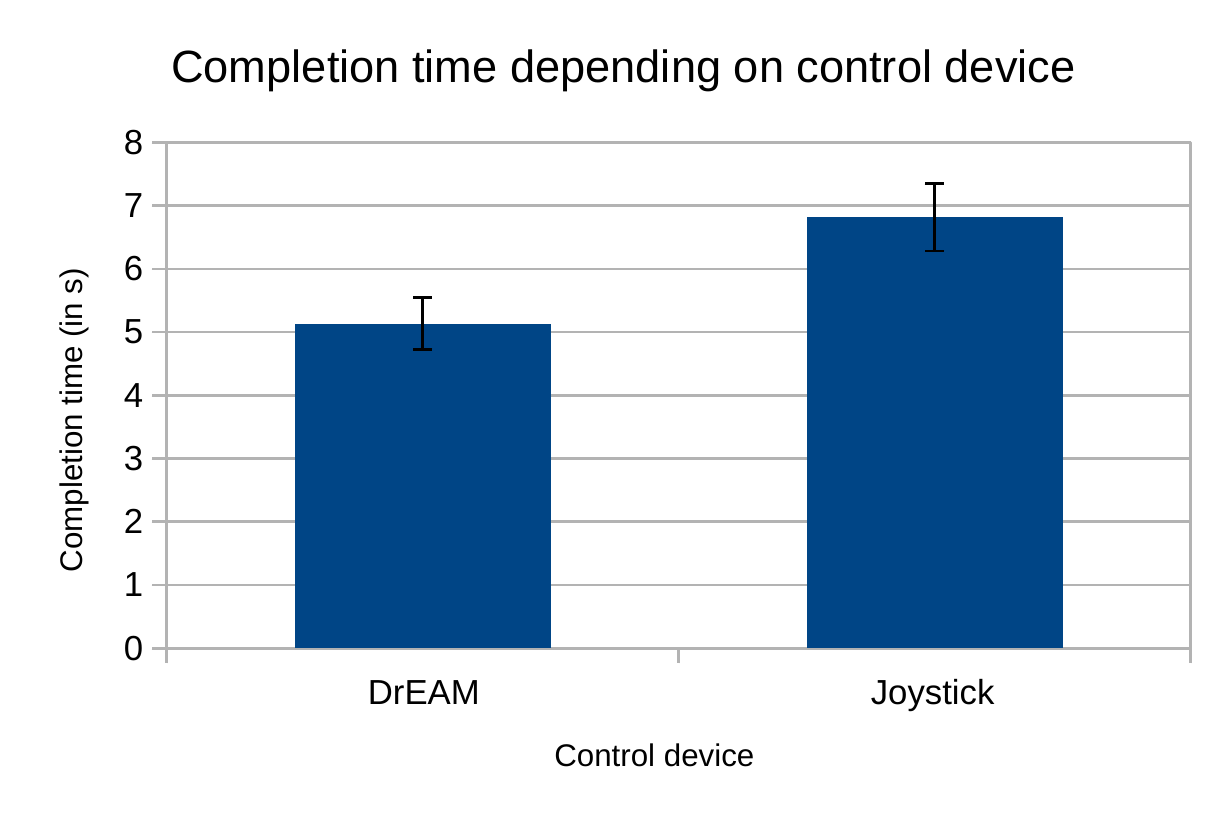}
		\caption{MCT for all journeys}
	\end{subfigure}

	\caption{Results of the flight logs analysis}
	\label{res-mesure}
\end{figure}

MLE, MCT and MYE are shown in figure \ref{res-mesure}. During the tests, 162 journeys were correctly performed by the pilots, 81 with a joystick and 81 with the immersive environment. Latency between immersive room and flight room was about 0.06s and no jitter has been registered, which could have disturbed flight. DrEAM was sending order at a 100Hz frequency and UAV was allway responding to these messages at the same frequency. \\

We obtain $MLE_{DrEAM}=0.104m$ for DrEAM and $MLE_{Joystick}=0.389m$ for the joystick. \\
MYE are $MYE_{DrEAM}=0.140$ for DrEAM and $MYE \psi_{Joystick}=0.252$ for the joystick. \\
MCT are $MCT_{DrEAM}=5.13s$ for DrEAM and $MCT_{Joystick}=6.81s$ for the joystick.

\section{Results Analysis}
Eight participants tested two control of UAV interactions, one with a joystick and one with an immersive environment. Their performances were registered by an application and processed after the experiment and they were given a TLX form after each experiment to evaluate their user experience.
Users generally had better performances with the immersive environment than with the joystick and felt less mental and temporal demand. This tends to confirm the initial hypothesis, which was that DrEAM would increase the handling ability, without precision loss, in comparison with a direct view control, in particular the composition of movements through degrees of freedom.

\subsection{Position control versus speed control}
User have a tendancy to think they perform better in the immersive environment according to the instructions we gave them, and this is coherent with the flight logs analysis.\\
MYE can be seen as the most important data from the flight information. It shows how hard it is to keep tracking a point while moving laterally. Here, it shows that the users follow better the given path with DrEAM.
This could be due to the fact that using a joystick forces the pilote to control speed (rotation speed, movement speed), using Dream he controls directly the yaw and the position with the immersive environment. This is corroborated by the meaningful lower MLE. Ease of placing one's own body at precise places may have increased the precision of DrEAM during the experiment. \\
Concerning the sensation of performance provided by DrEAM metaphor, user can see the UAV feedback while he manipulate the UAV itself, which helps to estimate the quality of the task.
When a user controls the UAV with the joystick, his hands, used to control the UAV, are not in his field of vision at the same time as the UAV, which can be a source of disturbance for unexperimented users.

\subsection{Gesture analysis}
\begin{figure}[t]
	\center
  \includegraphics[width=\linewidth]{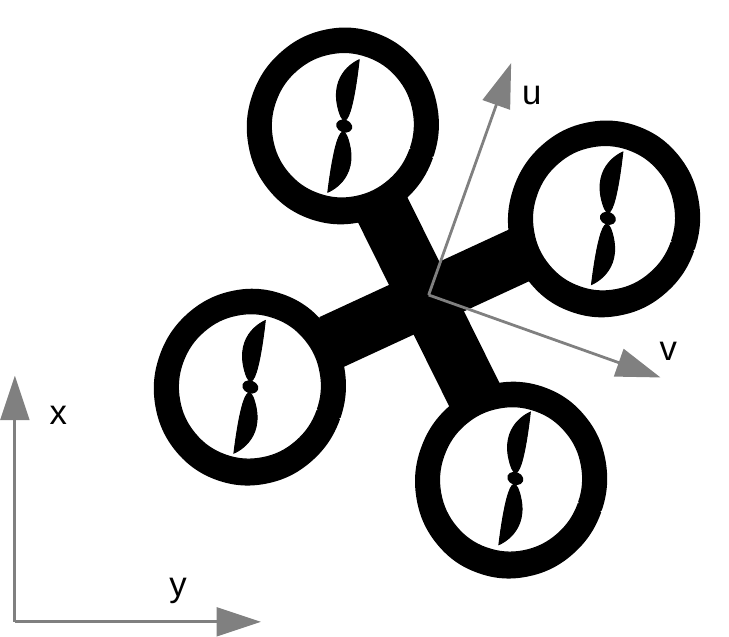}
	\caption{Frames used in control task. Inertial frame (in bottom left corner) has a fixed origin, body frame (on the UAV) has the center of the UAV as origin. $x$ and $y$ are fixed, $u$ and $v$ depend on the UAV orientation}
	\label{frames}
\end{figure}

Using the joysticks, user must activate two joysticks to perform both movements: placing the UAV at a specific place and orienting it. Furthermore, user can perceive the movement decomposition needed for this action. He is able to determine what gesture he performed to translate the UAV and what other gesture he performed to rotate it. \\
One the other hand, placing one's own hand at a specific place with a specific orientation can be performed using only one one simple gesture. In our task, using DrEAM, orient the UAV and place it at a specific place can be seen as \textit{one} simple task. 
User always has to perform a frame change when he controls with the joysticks, because he controls it in the body frame (UAV frame) however the task is given in the inertial frame (Pilote's frame) (see figure \ref{frames} for frames description). This can explain the high workload, because in the exocentric metaphor, user doesn't have to care about frames and orientation.
\the\linewidth

\subsection{Physical and Mental demand of the system}
We tried to avoid perspective effect to concentrate only on control performance and ease of control but it cannot be denied that users are biased when they try to align correctly the UAV with the target. However, some users did hit the cylinder in the flight zone with joystick and not with DrEAM: even with a physical marker at important places, some users were lost in joystick controls.
This means that this effects cannot be the lonely explanation for the difference of performance concerning lateral error, and hardness of control is probably the biggest reason.
Almost every subject beginning with the CAVE asked the experimenter why they had to perform a so simple task:  they considered the task as easy, sometime boring. In comparison, some users starting with the joystick asked us if it would be easier with DrEAM: they considerer the task hard. 
Users evaluated the immersive environment as more physically demanding which can be easily explained by the fact that the joystick does not need any arm or body movement to be used. Moreover, with experience settings, user needs to perform one sidestep to perform the task with DrEAM (we could not show any impact of this sidestep on navigation, but we will avoid such step in the future)
It was not possible to conclude with temporal demand and frustration, however, since it is possible to control the virtual UAV faster than the real UAV, user can release it's attention a short time and temporal demand should appear higher with joystick according to our hypothesis. \\
\\

However, all the results presented in this part have to be used very cautiously, they are based on a very little sample of novice users, and should only be seen as a first encouraging step in the study of exocentric metaphors for control of UAV.

\section{Conclusion and future works}

This study proposed a first step in the field of exocentric metaphors for UAV control.
According to the results of the TLX forms, an exocentric interaction seems to have better User Experience performances than joystick for control of UAV and further experiments should be lead with others devices.
The task was designed to compose two degrees of freedom which are yaw rotation and lateral translation. This could explain the difference of results in favor of DrEAM.\\
It shows the impact of control with an exocentric metaphor, in particular concerning composition of degrees of freedom, even for simple path.

\subsection{Experiment improvements}
We could have tested DrEAM against a joystick control within the Virtual Environment, but we can reasonably think that results would have been worse according to all criteria in this study than in direct view piloting with joystick.
However, there are a lot of parameters to study to produce optimized exocentric metaphors for control of UAV.
We used a CAVE-like environment  but further investigation could be lead with head mounted devices (HMD), despite the smaller embodiment provided by the HMD and the direct presence of one's hand in a CAVE-like environment.
Future work could compare this metaphor with other NUI in order to check the benefits of such an environment over intuitiveness and learning ability.

\subsection{Metaphor improvements}
Moreover, a lot of parameters can impact the performances of this metaphor and should be investigated : the size of the VUAV, of the PUAV, the scale of the environment, but also the command laws used or the speed of the real UAV. Impact of network issues could also be investigated.
Optimized moves in a WIM as widely explored in the literature and we should try a 2 handed interface to better divide the interaction witht the UAV and the interaction with environment parameters.
This study shows that some fatigue due to gesture will have to be considered. We are currently working on path planning to limit it, in order to let user pre-define a path and take back the direct control if the situation needs it. \\
Furthermore, a big limitation of such control interaction is that it needs a strong model of the environment, that's why it will be necessary to combine data from sensors with environment model and think about best method to perform such task.


\newpage
\bibliographystyle{refig-alpha}
\bibliography{bibliographie}

\end{document}